\documentstyle[preprint,aps]{revtex}
\tightenlines
\begin{document}

\title{\large \bf Tunneling and Chaos}

\author{S. Tomsovic}

\address{Department of Physics, Washington State University,
Pullman, WA99164-2814, USA\footnote{e-mail address: tomsovic@wsu.edu}}

\date{\today}

\maketitle

\begin{abstract}

Over the preceeding twenty years, the role of underlying classical
dynamics in quantum mechanical tunneling has received considerable
attention.  A number of new tunneling phenomena have been uncovered that
have been directly linked to the set of dynamical possibilities arising
in simple systems that contain at least some chaotic motion.  These
tunneling phenomena can be identified by their novel $\hbar$-dependencies
and/or statistical behaviors.  We summarize a sampling of these phenomena
and mention some applications.  

\end {abstract}


\section{Introduction}
\label{intro}

Quantum mechanical tunneling encompasses a multitude of fascinating 
effects that violate classically forbidden processes.  In spite of the
opposing quantum and classical behaviors, it is possible, through the use
of semiclassical methods, to construct a theory requiring only knowledge
of the classical dynamics.  For example, one of the most familiar
problems is that of potential barrier penetration.  For a one dimensional
barrier and supposing that coupling to the environment can be neglected,
a number of techniques are available that analytically continue the
classical motion either into a complexified phase space or through the
use of complexified time.  However, attempts to generalize these
techniques run into fundamental problems for systems with more degrees of
freedom, even without adding in the complication of environmental
coupling.  A theory of multidimensional tunneling has thus resisted any
single, universal treatment.  

It turns out that the difficulties can be traced back to the broad variety
of underlying classical dynamics that emerge from multidimensional
systems, and which are absent from one-degree-of-freedom systems.  With
better identification of the difficulties has come a number of new and
novel tunneling phenomena linked to the presence of at least some chaotic
motion.  An excellent and fairly recent overview of the subject has been
published as a pedagogical course by Creagh~\cite{creagh}.  

As chaos is ubiquitous in the dynamics of simple systems, i.e. those
possessing a few degrees of freedom, there has been a major shift in
focus towards investigating non-integrable dynamics.  Whereas
integrability may be {\it ``pleasing''} in that it is amenable to
analytic analysis, i.e.~action-angle variables exist that render the
dynamics trivial, in some sense it is extremely rare or can be considered
as special.  A few other categories of dynamics need to be introduced. 
For systems regarded as {\it near-integrable}, the dynamical instability
tends to be quite weak, and perturbation theories often apply.  Systems
possessing a globalized chaotic region in phase space and some stable
motion are regarded as having a {\it mixed phase space} with a
moderate level of instability.  The extreme limit of {\it fully chaotic} 
dynamical systems often comes with highly unstable motion.  Each of these
dynamical possibilities creates the opportunity for unique tunneling
behaviors.  

After a short description of a couple essential concepts, the rest of the
paper gives a cursory outline of several advances and organizes them
loosely by dynamical system type.  No attempt has been made to be
complete.  It is just intended to give the flavor of the new dynamical
directions under study.

\section{Preliminaries}
\label{prelim}

With a few exceptions, we restrict our discussion mainly to the spectral
properties of simple, environmentally isolated, bounded, few
degrees-of-freedom (D) systems.  In fact, paradigms of $2$-D and 
periodically kicked $1$-D systems have been invoked in the bulk of the
recent work.  The nature of eigenstates or the time evolution of
initially localized states is an extremely interesting subject which
ought to be included in such a discussion as this, but shall not be due
to space constraints.  More complete discourse and references 
can be found in Ref.~\cite{creagh}.

\subsection{Integrable Systems}
\label{integ}

An inherent signature of tunneling is the existence of almost
degenerate multiplets in a spectrum.  For systems with a single
reflection symmetry, doublets would appear if some positive measure of
the classical trajectories occupied distinct phase space regions from
their reflection symmetric counterparts.  A semiclassical quantization of
the motion and its symmetric partner would lead to the expectation of a
degeneracy.  The energy splitting of the degeneracy, $\Delta E$, would be
a measure of the strength of the tunneling.  For integrable systems,
semiclassical formulae take the form

\begin{equation}
\label{dele}
\Delta E = A({\bf J})\hbar\exp\left(-S({\bf J})/\hbar\right)
\end{equation}
where ${\bf J}$ denotes the action variables of the system, and $\{A({\bf
J}),S({\bf J})\}$ are smooth scalar functions of their arguments. 
Typically, $A({\bf J})$ is closely related to angular frequencies, and
$S({\bf J})$ to the action accumulated over a particular cyclical,
classically forbidden path defined within the complexified dynamics.  

Of particular interest is the $\hbar$-dependencies of both the amplitude
and the argument of the exponential.  Often it is possible to control a
parameter, such as a wavevector, which effectively allows a tuning of the
value of $\hbar$.  In a log plot versus $\hbar$, the tunneling splitting
should exhibit a smooth, linear slope.  The prefactor shows up as a
logarithmic correction, and may be rather difficult to establish, but
leaves the $\hbar$-dependence smooth.  It is the contrast to this
behavior that sets the new phenomena apart.  

\subsection{Dynamical Tunneling}
\label{dynam}

Many of the recently discovered tunneling phenomena involve a counterpart
to potential barrier penetration dubbed dynamical tunneling by Davis and
Heller~\cite{davis}.  It frequently arises in the models of the nuclear
motions of small molecules~\cite{example}.  In dynamical tunneling, there
is no energetic or potential barrier.  Rather, the classically forbidden
motions are dynamical in nature.  In other words, constants of the motion
other than energy are responsible for the forbidden features.  The view
from the underlying phase space structure can look indistinguishable
between the two tunneling mechanisms.  Separatrices, to which a potential
barrier would give rise in phase space, or resonances can be responsible
for isolating the symmetrically related motions.  Unstable periodic orbits
with their respective homoclinic tangles can also provide the barrier for
dynamical tunneling.  Curiously though, despite the phase space
similarities, the spectral behavior can appear to vary in opposing
directions.  The potential barrier case leads to increasing tunneling
splittings as energy is increased, and beyond the peak barrier energy the
tunneling splittings vanish in the spectrum.  On the other hand, many
dynamical systems retain dynamical tunneling multiplets to infinite
excitation energy with many of the splittings approaching zero in the
limit.  For example, homogeneous potentials lead to situations in which a
constant fraction of the levels are involved with tunneling at all
energies and the local energy average splitting decreases with
increasing energy.  

\subsection{Chaotic eigenstates}
\label{chasta}

For integrable systems, semiclassical theory gives a one-to-one
correspondence between particular classical motion and the eigenstates. 
Through the D relations

\begin{equation}
J_i = 2\pi\hbar \left(n_i + {\nu_i \over 2}\right)
\end{equation}
each energy level and eigenstate is related to a specific set of
classical actions, $\{{\bf J}\}$.  Topologically speaking, all the
eigenstates are similar and can be considered to belong to a regular
class.  This association fundamentally breaks down once chaotic
motion exists in the system~\cite{ein}.  For near-integrable and mixed
phase space systems a non-neglible fraction of the phase space continues
to be covered by motion possessing good action variables locally.  A
roughly corresponding fraction of the eigenstates continues to satisfy
the criteria for being regular~\cite{percival,btu}.  

The remainder of the eigenstates can be lumped into a single irregular
or chaotic class defined as being other than regular, but this
underemphasizes the range of eigenstate characteristics that become
possible once released from the burden of having to reflect regular
motion.  We mention three important examples of distinct irregular
behaviors;  this is not an exhaustive scheme.  Crudely speaking, the {\it
ergodic} subclass behaves locally like random waves subject to the
ergodic measure
$\delta(E-H(p,q))$ as it applies to wavefunctions~\cite{berry}.  If a
system is not fully chaotic, it is understood that the $(p,q)$
coordinates are restricted to a subset of points approaching arbitrarily
closely to a single chaotic trajectory.  For this class to exist, the
classical exploration of the ergodic subset of phase points must take
place on a time scale short compared to the Heisenberg time scale that
derives from the mean quantum level spacing.  A second possibility arises
because it is often possible to identify a variable in which the dynamics
is diffusive.  In this case, the irregular class of eigenstates may be
strongly {\it localized} in the Anderson sense~\cite{anderson}.  The
kicked rotor provides the classic example~\cite{fishman}.  A
third class has been dubbed as being {\it
hierarchical}~\cite{ketzmerick,btu}.  They are associated with the
hierarchy of time scales existing in the dynamics found at the
boundaries between regular and chaotic motion.  The states exhibit
fractal character over some range of scales cutoff by $\hbar$ at one end,
and the nature of the regular-chaos boundary on the other.  

\section{Near-integrable systems}
\label{near}

\subsection{An integrable approximation}
\label{integrable}

Our first example of a tunneling phenomenon requiring a serious look at
the underlying dynamics is provided by a work of Ozorio de
Almeida~\cite{ozorio}, and is exceptional in that it is the only topic
discussed here in which the chaos is explicitly removed before treating
the problem.  However, it gives a theoretical approach to understanding
dynamical tunneling in the presence of resonances, as are created by
generic perturbations, and so it has been included.  For small
perturbations of an integrable system, motion in the neighborhood of low
order rational ratios of angular frequency are replaced by resonances. 
The resonances may be enveloped by narrowly confined chaotic zones,
however if the perturbation is weak enough, it is possible to average the
Hamiltonian over the rapidly evolving angle variable of its unperturbed
form.  Effectively, through the combination of averaging and canonical
transformations, the perturbed system is replaced by an integrable system
whose phase space structure appears locally as a distorted pendulum.  Two
different tunneling processes emerge.  The more familiar process is the
tunneling between motion above and below the resonance leading to avoided
crossings as a system parameter is varied.  The other is a sort of
tunneling of the state unto itself.  If the resonance is large enough to
quantize, its energy will be shifted from the standard semiclassical
quantization because the wavefunction has ``tails'' that communicate
directly and weakly from any lobe of the resonance to its neighboring
lobes.  

\subsection{Analytic continuation}

It is not always necessary nor correct to average away the weak chaos
present in the dynamics, and a new tunneling dependence on $\hbar$ was
derived in the work of Wilkinson~\cite{wilkinson}.  He was able to
evaluate, by saddle point methods, the tunneling splitting integrals with
the help of an analytic continuation of the regular dynamical motion. 
The explicit expressions contain a different prefactor power of $\hbar$

\begin{equation}
\label{delp}
\Delta E = \sum_k A_k({\bf J})\hbar^{3/2}\exp\left(-S_k({\bf
J})/\hbar\right)
\end{equation}
than found in Eq.~(\ref{dele}), and multiple contributions are possible. 
In this case, the saddle points are isolated in distinction to the contour
integrations necessary in integrable systems.   Although, the original
derivation required separation of the motions in configuration space, the
canonical form suggested, and he conjectured, that the results apply to
more general dynamical tunneling situations.   A criterium was also given
for the cross-over between the regimes of applicability of
Eqs.~(\ref{dele}) and (\ref{delp}).  With increasing perturbation, the
analytic continuations can become limited to such an extent that the
saddle points cease to exist, and the procedure eventually breaks
down.  
 
\section{Mixed phase space systems}
\label{mixed}
We turn now to tunneling phenomena where chaos is taking a direct role in
the process.  

\subsection{Chaos-assisted tunneling}
\label{cat}

Consider a mixed phase space system whose parameters are such that each
of its eigenstates fall mainly into a regular or ergodic class.  Suppose
further that there exists reflection symmetry leading to duplicate copies
of regular trajectories that are well separated.  One anticipates
semiclassically degenerate doublets in the spectrum in proportion to the
measure of regular motion.  It turns out that these doublets are split by
tunneling mechanisms completely different than discussed in
Sect.~\ref{near}.  Instead, the tunneling proceeds indirectly
via a compound process of wave amplitude breaking off bit by bit from the
initial state near one regular domain, transporting chaotically to the
neighborhood of its symmetric regular domain, and there reassembling its
reflection of the initial state~\cite{tu,btu}.  The fundamental process,
and unsolved theoretical problem, is how the ``phase space'' tails of
regular and ergodic wavefunctions join smoothly.  The involvement of the
ergodic eigenstates and spectrum create tunneling splitting fluctuations
over several orders of magnitude~\cite{tu,btu,ullmo} which overshadow any
mean behavior similar to Eq.~(\ref{dele}).  Random matrix models have
been introduced and shown to predict the observed splitting fluctuations
successfully~\cite{tu}. 

The first experimental evidence for chaos-assisted tunneling in a
microwave cavity version of the annular billiard introduced by Bohigas et
al.~\cite{boh} has recently been reported by Dembowski et
al.~\cite{dembo}.  Note that it had previously been argued that
chaos-assisted tunneling is involved in the decay, to normal deformation,
of superdeformed nuclear states~\cite{aberg}, but the decay involves
many-body physics that makes it a less straightforward interpretation. 
It nevertheless involves the coupling of regular and presumably ergodic
states.

Several other important works have appeared, especially with regards to
kicked systems or quantum maps~\cite{lin}.  However, we briefly discuss
just one more issue~\cite{tomso}.  The question arises whether symmetry is
necessary; suppose that in a particular experimental situation it cannot
be exactly maintained.  Is there a possibility of seeing and confirming
that chaos-assisted tunneling is occuring?  In fact, symmetry is not
required at all if two parameters can be controlled and independently
varied.  One parameter must force crossings between regular states that
the tunneling will be splitting.  The other parameter shifts the
irregular levels with respect to the crossing levels.  In direct
tunneling, the tunneling splitting at the crossing is not influenced by
the presence of other levels.  In chaos-assisted tunneling, the splitting
is completely determined by the whereabouts of the irregular levels, and
the magnitudes of their avoided crossings with the regular levels.  In
the avoided crossings' parametric dependence, these distinctions are
unmistakable.

\subsection{Chaos-assisted ionization}
\label{cai}

Chaos-assisted tunneling also arises in systems that are coupled to a
continuum~\cite{zakzrewski,nockel}.  In the first reference, the authors
have investigated certain regimes of the hydrogen atom exposed to either
linearly or circularly polarized microwaves where nondispersive electronic
wave packets can be created.  There regular classical motion exists
embbedded in chaotic dynamical regions of phase space.  Trajectories
diffusively transport to ionization thresholds.  It happens that the
tails of the electronic wave packets are connected by quantum tunneling
with ergodic states localized in the chaotic dynamical part of the phase
space which are in turn connected to the continuum.  The authors
found peculiar fluctuations of energies and ionization
rates and successfully modeled them with the same basic random matrix
model approach as introduced in~\cite{tu} extended to include continuum
coupling.  

In the second above reference, the authors investigate a quite different
physical system, micro-optical cavities.  These slightly deformed
circular cavities have long-lived whispering gallery resonances created
by total internal reflection.  Qualitatively, certain parameter regimes of
the system's asymptotic ray limit are dynamically similar to the above
nondispersing wave packets, i.e.~mixed phase space dynamics.  The
authors~\cite{nockel} find a range of deformations where it appears
that chaos-assisted tunneling must be invoked to explain larger than
expected lifetimes for these meta-stable resonances.  

\section{Fully chaotic systems}
\label{chaos}

For fully chaotic systems, recent works address tunneling between two
ergodic states or between two localized states.  We also give one last
example reminiscent of Ozorio de Almeida's tunneling of a state unto
itself\cite{ozorio} summarized in Sect.~\ref{integrable}; although instead
of involving a regular state, it involves an ergodic state.

\subsection{Tunneling between chaotic states}
\label{states}

In a series of works, Creagh and Whelan have studied the
tunneling behavior of ergodic states separated by a potential
barrier~\cite{cw}.  They began by introducing a 2D double well for which
the motion within each well was chaotic, and by giving a new approach to
generating a trace formula directly for the splittings.  A mean and
fluctuating behavior could be separated and indentified with specific
complex orbits.  The fluctuations were controlled by orbits homoclinic to
the real orbit connected to the optimal tunneling path through the
barrier in a very similar way as a homoclinic orbit sum controls the
autocorrelation function of chaotic wave packets~\cite{tomso2}.  They
also studied the statistical properties of the tunneling splittings which
are given by a generalized Porter-Thomas distribution.  

\subsection{Tunneling between strongly localized states}
\label{localized}

Strong localization provides a quantum dynamical barrier that separates
states as opposed to the classical dynamical barriers originally
invoked in the definition of dynamical tunneling.  For systems
possessing an appropriate symmetry, multiplets of states that are
further than a localization length from the ``symmetry lines'' will exist
whose degeneracy is broken by tunneling~\cite{iz}.  With a generalization
of the kicked rotor~\cite{blumel}, Casati et al.~\cite{iz} found, not
surprisingly, that pairs of ``double-humped'' states existed, and that
their quasi-energies were split similar to the expectation for Mott
states,
\begin{equation}
\Delta \theta \sim \exp( -c_0 n_0/ l)
\end{equation}
where $l$ is the localization length, $n_0$ the number of states away
from the symmetry line, and $c_0$ a constant.  The pairs did not exist
too close to the momentum origin, and the localization length for
states localized near the origin appeared to be a factor two different in
their localization lengths.  

\subsection{Tunneling influence on band structure}
\label{band}

The band structure of a periodic system that happens to be completely
chaotic classically can be heavily influenced by tunneling.  For strong
enough kicking strength, the kicked Harper equation serves as an
excellent example~\cite{pl}.  Ergodic trajectories diffuse through the
lattice.  However, classically forbidden complex trajectories may hasten
the process.  LeBoeuf and Mouchet~\cite{leboeuf} have treated precisely
this problem.  As a function of kicking strength near integer values,
allowed trajectories pass from the real to complex domains.  With the use
of the Gutzwiller trace formula in relations deriving from the spectral
determinant, they compared results with and without the inclusion of
these complex trajectories.  They found that in some cases the inclusion
of the complex trajectories significantly altered the spectrum as a
function of the Bloch angle (quasi-momentum), and thus had an important
impact on the band structure.  

\section{Summary}

The past ten or fifteen years have taught us that there exists a vastly
richer collection of manifestations of tunneling than provided by the one
degree-of-freedom double well.  Despite several beautiful works, many,
very basic, theoretical problems remain such as developing a theory for
how regular and ergodic wavefunctions couple.  In addition, new physical
applications continue to surface.  Given that tunneling impacts all
domains of wave physics, not just quantum mechanics, the future of the
subject promises to be very rewarding. 

\subsection*{Acknowledgments}

We gratefully acknowledge the efforts of the organizers, S.~Aberg,
K.-F.~Berggren, and P.~Omling, of the Nobel Symposium entitled ``Quantum
Chaos Y2K'' as well as the support of the Nobel Foundation.  Our
apologies to S.~Adachi, K.~Ikeda, and A.~Shudo for not including their
tunneling studies due to space constraints; see~\cite{creagh} for a
summary and references to their works.

\end{document}